\documentclass[preprint,prd,aps,showpacs,showkeys,nofootinbib]{revtex4}
\usepackage{graphicx}
\pdfoutput=1
\usepackage{hyperref}
\begin{document}


\title{Higgs boson decay $h\rightarrow Z\gamma$ and muon magnetic dipole moment  in the $\mu\nu$SSM}

\author{Chang-Xin Liu$^{a,b}$\footnote{LIUchangxinZ@163.com},
Hai-Bin Zhang$^{a,b}$\footnote{hbzhang@hbu.edu.cn},
Jin-Lei Yang$^{a,b,c}$\footnote{JLYangJL@163.com},
Shu-Min Zhao$^{a,b}$\footnote{zhaosm@hbu.edu.cn},
Yu-Bin Liu$^{d}$\footnote{liuyb@nankai.edu.cn},
Tai-Fu Feng$^{a,b,e}$\footnote{fengtf@hbu.edu.cn}}

\affiliation{$^a$Department of Physics, Hebei University, Baoding, 071002, China\\
$^b$Key Laboratory of High-precision Computation and Application of Quantum Field Theory of Hebei Province, Baoding, 071002, China\\
$^c$Institute of theoretical Physics, Chinese Academy of Sciences, Beijing, 100190, China\\
$^d$School of Physics, Nankai University, Tianjin 300071, China\\
$^e$College of Physics, Chongqing University, Chongqing, 400044, China}

\begin{abstract}
To solve the $\mu$ problem and generate three tiny neutrino masses in the MSSM, the $\mu$ from $\nu$ Supersymmetric Standard Model ($\mu\nu$SSM) introduces three singlet right-handed neutrino superfields, which lead to the mixing of the Higgs doublets with the sneutrinos. The mixing affects the lightest Higgs boson mass and the Higgs couplings. The present observed 95\% CL upper limit on signal strength of the 125 GeV Higgs boson decay $h\rightarrow Z\gamma$ is 6.6, which still is plenty of space to prove the existence of new physics. In this work, we investigate the signal strength of the 125 GeV Higgs boson decay channel $h\rightarrow Z\gamma$ in the $\mu\nu$SSM. Besides, we consider the two-loop electroweak corrections of muon anomalous magnetic dipole moment (MDM) in the model, which also make important contributions compared with one-loop electroweak corrections.
\end{abstract}


\maketitle

\section{Introduction\label{sec1}}

A great success of the Large Hadron Collider (LHC) is the discovery of the Higgs boson \cite{mh-ATLAS,mh-CMS}. Combining the updated data~\cite{mh-LHC,mh-CMS1,mh-ATLAS1}, the measured mass of the Higgs boson now is~\cite{PDG1}
\begin{eqnarray}
m_h=125.10\pm 0.14\: {\rm{GeV}}.
\label{mh-exp}
\end{eqnarray}
Therefore, the accurate Higgs boson mass gives the most stringent constraint on parameter space for the standard model and its various extensions. The next step is focusing on searching for the properties of the Higgs boson. Now, the signal strengths for the Higgs boson decays $h\rightarrow\gamma\gamma$, $h\rightarrow VV^*$ ($V=Z,W$) and $h\rightarrow f\bar{f}$ ($f=b,\tau$) can be detected by precise values. The signal strength for their combined final states is $1.10\pm0.11$~\cite{PDG1}, which is consistent with the value of the standard model (SM) in the error range. The LHC also has reported the searches for the rare decay process $h\rightarrow Z\gamma$ \cite{hZr-CMS1,hZr-ATLAS1,hZr-ATLAS2,hZr-CMS2}. But no evidence for the decay $h\rightarrow Z\gamma$ is observed and the present observed 95\% CL upper limit on its signal strength is 6.6~\cite{hZr-ATLAS2}. So, for the decay $h\rightarrow Z\gamma$, there is still plenty of space for new physics (NP). In this paper, we will investigate the decay  $h\rightarrow Z\gamma$ in a new physics model to show how large new physics contributions. In the future, high luminosity or high energy large collider~\cite{ref-100pp,ref-HL,ref-CEPC} will detect the Higgs boson decay $h\rightarrow Z\gamma$, which may see the indication of new physics. Moreover, the measurement of $h\rightarrow Z\gamma$ and its rate compared to $h\rightarrow\gamma\gamma$ is crucial for broadening our understanding of the electroweak symmetry spontaneously broken (EWSB) pattern~\cite{hZr-5,Higgs-SM,Higgs}. Testing the SM nature of the Higgs boson state and inspecting possible deviations in its coupling to the SM particles will represent a major undertaking of modern particle physics and a probable probe of new physics.
Within various theoretical frameworks, the 125 GeV Higgs boson decay $h\rightarrow Z\gamma$ has been discussed
\cite{hZr-5,Higgs-SM,Higgs,hZr-1,hZr-2,hZr-3,hZr-4,hZr-4-1,hZr-6,hZr-6-1,hZr-7,hZr-8,hZr-9,hZr-10,
hZr-11,hZr-12,hZr-13,hZr-14,hZr-14-1,hZr-15,hZr-16,hZr-17,hZr-18,hZr-19,hZr-20,
hZr-21,hZr-22,hZr-23,hZr-24,hZr-25,hZr-26,hZr-27,hZr-28,hZr-29}.

As one of the candidates of new physics, the $\mu$ from $\nu$ supersymmetric standard model ($\mu$$\nu$SSM) \cite{mnSSM,mnSSM1,mnSSM1-1,mnSSM2,mnSSM2-1,Zhang1,Zhang2} can solve the $\mu$ problem~\cite{m-problem} of the minimal supersymmetric standard model (MSSM)~\cite{MSSM,MSSM1,MSSM2,MSSM3,MSSM4} through introducing three singlet right-handed neutrino superfields $\hat{\nu}_i^c$ ($i=1,2,3$). The neutrino superfields lead the mixing of the neutral components of the Higgs doublets with the sneutrinos, which is different from the Higgs sector of the MSSM. In our previous work, the Higgs boson decay modes $h\rightarrow\gamma\gamma$, $h\rightarrow VV^*$ ($V=Z,W$), $h\rightarrow f\bar{f}$ ($f=b,\tau$), $h\rightarrow \mu\tau$, and the masses of the Higgs bosons in the $\mu\nu$SSM have been researched~\cite{HZrr,muon,MASS}. In this paper, we will investigate the 125 GeV Higgs boson decay channel $h\rightarrow Z\gamma$ in the $\mu\nu$SSM to see how large new physics contributions.

In addition, the current difference between the experimental measurement \cite{muon-exp} and SM theoretical prediction of the muon anomalous magnetic dipole moment (MDM) \cite{PDG1},
\begin{eqnarray}
\Delta a_{\mu}=a_{\mu}^{exp}-a_{\mu}^{SM}=(26.8\pm7.7)\times10^{-10},
\label{MDM-exp}
\end{eqnarray}
represents an interesting but not yet conclusive discrepancy of 3.5 standard deviation, which still stands as a potential indication of the existence of new physics. Up to now, several predictions for the muon anomalous MDM have been discussed in the framework of various SM extensions
\cite{Abel:1991dv,Moroi:1995yh,Feng:2001tr,Martin:2001st,Arnowitt:2001,Diaz:2002tp,Feng:2006,Feng:2008cn,
Feng:2008nm,Feng:2009gn,Yang:2009zzh,Cheung:2009fc,Zhao:2014dxa,Conto:2017,Queiroz:2018,muon-a,muon-D}.
In near future, the Muon g-2 experiment E989 at Fermilab~\cite{ref-muon-exp,ref-muon-exp1} will measure the muon anomalous MDM with unprecedented precision, which may reach a 5$\sigma$ deviation from the SM, constituting an augury for new physics. In our previous work, we have studied the muon MDM at one-loop level in the $\mu\nu$SSM~\cite{muon}. To be more precise, here we will consider the two-loop diagrams of the muon anomalous MDM in the framework of the $\mu\nu$SSM. Simultaneously, the accurate theoretical prediction of the muon anomalous MDM can conduce to constrain strictly the parameter space of the model.

The paper is organized as follows. In Sec.~\ref{sec2}, we introduce the $\mu\nu$SSM briefly, about the superpotential and the soft SUSY-breaking terms. In Sec.~\ref{sec3}, we give the decay width and the signal strength of $h\rightarrow Z\gamma$. Sec.~\ref{sec4} includes the two-loop electroweak  corrections of the muon anomalous MDM. Sec.~\ref{sec5} and Sec.~\ref{sec6} respectively show the numerical analysis and summary. Some formulae are collected in Appendix.

\section{the $\mu\nu$SSM\label{sec2}}

In addition to the MSSM Yukawa couplings for quarks and charged leptons, the superpotential of the $\mu\nu$SSM contains Yukawa couplings for neutrinos, two additional types of terms involving the Higgs doublet superfields $\hat H_u$ and $\hat H_d$, and the right-handed neutrino superfields  $\hat{\nu}_i^c$,~\cite{mnSSM}
\begin{eqnarray}
&&W={\epsilon _{ab}}\left( {Y_{{u_{ij}}}}\hat H_u^b\hat Q_i^a\hat u_j^c + {Y_{{d_{ij}}}}\hat H_d^a\hat Q_i^b\hat d_j^c
+ {Y_{{e_{ij}}}}\hat H_d^a\hat L_i^b\hat e_j^c \right)  \nonumber\\
&&\hspace{0.95cm}
+ {\epsilon _{ab}}{Y_{{\nu _{ij}}}}\hat H_u^b\hat L_i^a\hat \nu _j^c -  {\epsilon _{ab}}{\lambda _i}\hat \nu _i^c\hat H_d^a\hat H_u^b + \frac{1}{3}{\kappa _{ijk}}\hat \nu _i^c\hat \nu _j^c\hat \nu _k^c ,
\label{eq-W}
\end{eqnarray}
where $\hat H_u^T = \Big( {\hat H_u^ + ,\hat H_u^0} \Big)$, $\hat H_d^T = \Big( {\hat H_d^0,\hat H_d^ - } \Big)$, $\hat Q_i^T = \Big( {{{\hat u}_i},{{\hat d}_i}} \Big)$, $\hat L_i^T = \Big( {{{\hat \nu}_i},{{\hat e}_i}} \Big)$ (the index $T$ denotes the transposition) represent $SU(2)$ doublet superfields, and $\hat u_i^c$, $\hat d_i^c$, and $\hat e_i^c$ are the singlet up-type quark, down-type quark and charged lepton superfields, respectively.  In addition, $Y_{u,d,e,\nu}$, $\lambda$, and $\kappa$ are dimensionless matrices, a vector, and a totally symmetric tensor.  $a,b=1,2$ are SU(2) indices with antisymmetric tensor $\epsilon_{12}=1$, and $i,j,k=1,2,3$ are generation indices. The summation convention is implied on repeating indices in the following.

In the superpotential, if the scalar potential is such that nonzero vacuum expectation
values (VEVs) of the scalar components ($\tilde \nu _i^c$) of the singlet neutrino superfields $\hat{\nu}_i^c$ are induced, the effective bilinear terms $\epsilon _{ab} \varepsilon_i \hat H_u^b\hat L_i^a$ and $\epsilon _{ab} \mu \hat H_d^a\hat H_u^b$ are generated, with $\varepsilon_i= Y_{\nu _{ij}} \left\langle {\tilde \nu _j^c} \right\rangle$ and $\mu  = {\lambda _i}\left\langle {\tilde \nu _i^c} \right\rangle$,  once the electroweak symmetry is broken. The last term generates the effective Majorana masses for neutrinos at the electroweak scale. Therefore, the $\mu\nu$SSM can generate three tiny neutrino masses at the tree level through TeV scale seesaw mechanism
\cite{mnSSM1,neutrino-mass,neu-mass1,neu-mass2,neu-mass3,neu-mass4,neu-mass5,neu-mass6}.

In supersymmetric (SUSY) extensions of the standard model, the R-parity of a particle is defined as $R = (-1)^{L+3B+2S}$~\cite{MSSM,MSSM1,MSSM2,MSSM3,MSSM4}. R-parity is violated if either the baryon number ($B$) or lepton number ($L$) is not conserved, where $S$ denotes the spin of the concerned component field. The last two terms in Eq.~(\ref{eq-W}) explicitly violate lepton number and R-parity. R-parity breaking implies that the lightest supersymmetric particle (LSP) is no longer stable. In this context, the neutralino or the sneutrino are no longer candidates for the dark matter (DM). However, other SUSY particles such as the gravitino or the axino can still be used as candidates~\cite{mnSSM1,mnSSM1-1,neu-mass3,DM1,DM2,DM3,DM4,DM5,DM6}.

The dark matter candidate must be stable on the cosmic timescale, so that it is still around today \cite{Yangbinglin}. In Refs.~\cite{DM1,DM2,DM3,DM4}, the authors analyzed the gravitino dark matter candidate in the $\mu\nu$SSM,  whose lifetime is long lived compared to the current age of the Universe. The gravitino turns out to be an interesting candidate for DM, which may be searched through gamma-ray observations with Fermi-LAT. Recently, the axino dark matter candidate in the $\mu\nu$SSM also was analyzed~\cite{DM5,DM6}.

The general soft SUSY-breaking terms of the $\mu\nu$SSM are given by
\begin{eqnarray}
&&- \mathcal{L}_{soft}=m_{{{\tilde Q}_{ij}}}^{\rm{2}}\tilde Q{_i^{a\ast}}\tilde Q_j^a
+ m_{\tilde u_{ij}^c}^{\rm{2}}\tilde u{_i^{c\ast}}\tilde u_j^c + m_{\tilde d_{ij}^c}^2\tilde d{_i^{c\ast}}\tilde d_j^c
+ m_{{{\tilde L}_{ij}}}^2\tilde L_i^{a\ast}\tilde L_j^a  \nonumber\\
&&\hspace{1.7cm} +  m_{\tilde e_{ij}^c}^2\tilde e{_i^{c\ast}}\tilde e_j^c + m_{{H_d}}^{\rm{2}} H_d^{a\ast} H_d^a
+ m_{{H_u}}^2H{_u^{a\ast}}H_u^a + m_{\tilde \nu_{ij}^c}^2\tilde \nu{_i^{c\ast}}\tilde \nu_j^c \nonumber\\
&&\hspace{1.7cm}  +  \epsilon_{ab}{\left[{{({A_u}{Y_u})}_{ij}}H_u^b\tilde Q_i^a\tilde u_j^c
+ {{({A_d}{Y_d})}_{ij}}H_d^a\tilde Q_i^b\tilde d_j^c + {{({A_e}{Y_e})}_{ij}}H_d^a\tilde L_i^b\tilde e_j^c + {\rm{H.c.}} \right]} \nonumber\\
&&\hspace{1.7cm}  + \left[ {\epsilon _{ab}}{{({A_\nu}{Y_\nu})}_{ij}}H_u^b\tilde L_i^a\tilde \nu_j^c
- {\epsilon _{ab}}{{({A_\lambda }\lambda )}_i}\tilde \nu_i^c H_d^a H_u^b
+ \frac{1}{3}{{({A_\kappa }\kappa )}_{ijk}}\tilde \nu_i^c\tilde \nu_j^c\tilde \nu_k^c + {\rm{H.c.}} \right] \nonumber\\
&&\hspace{1.7cm}  -  \frac{1}{2}\left({M_3}{{\tilde \lambda }_3}{{\tilde \lambda }_3}
+ {M_2}{{\tilde \lambda }_2}{{\tilde \lambda }_2} + {M_1}{{\tilde \lambda }_1}{{\tilde \lambda }_1} + {\rm{H.c.}} \right).
\end{eqnarray}
Here, the first two lines contain mass squared terms of squarks, sleptons, and Higgses. The next two lines consist of the trilinear scalar couplings. In the last line, $M_3$, $M_2$, and $M_1$ denote Majorana masses corresponding to $SU(3)$, $SU(2)$, and $U(1)$ gauginos $\hat{\lambda}_3$, $\hat{\lambda}_2$, and $\hat{\lambda}_1$, respectively. In addition to the terms from $\mathcal{L}_{soft}$, the tree-level scalar potential receives the usual $D$- and $F$-term contributions~\cite{mnSSM1,mnSSM1-1}.

Once the electroweak symmetry is spontaneously broken, the neutral scalars develop in general the VEVs:
\begin{eqnarray}
\langle H_d^0 \rangle = \upsilon_d , \qquad \langle H_u^0 \rangle = \upsilon_u , \qquad
\langle \tilde \nu_i \rangle = \upsilon_{\nu_i} , \qquad \langle \tilde \nu_i^c \rangle = \upsilon_{\nu_i^c} .
\end{eqnarray}
One can define the neutral scalars as
\begin{eqnarray}
&&H_d^0=\frac{h_d + i P_d}{\sqrt{2}} + \upsilon_d, \qquad\; \tilde \nu_i = \frac{(\tilde \nu_i)^\Re + i (\tilde \nu_i)^\Im}{\sqrt{2}} + \upsilon_{\nu_i},  \nonumber\\
&&H_u^0=\frac{h_u + i P_u}{\sqrt{2}} + \upsilon_u, \qquad \tilde \nu_i^c = \frac{(\tilde \nu_i^c)^\Re + i (\tilde \nu_i^c)^\Im}{\sqrt{2}} + \upsilon_{\nu_i^c},
\end{eqnarray}
and
\begin{eqnarray}
\tan\beta={\upsilon_u\over\upsilon_d}.
\end{eqnarray}

In the $\mu\nu$SSM, the left- and right-handed sneutrino VEVs lead to the mixing of the neutral components of the Higgs doublets with the sneutrinos producing an $8\times8$ CP-even neutral scalar mass matrix, which can be seen in Refs.~\cite{mnSSM1,mnSSM1-1,Zhang1}. The mixing gives a rich phenomenology in the Higgs sector of  the $\mu\nu$SSM. In the large $m_A$ limit, we give an approximate expression for the lightest Higgs boson mass \cite{MASS},
\begin{eqnarray}
m_h^2 \approx  \xi_h  m_{H_1}^2,
\label{mh-app1}
\end{eqnarray}
where
\begin{eqnarray}
&&m_{H_1}^2\simeq m_{Z}^2 \cos{2\beta}^2+{2\lambda_i \lambda_i s_{W}^2c_{W}^2\over e^2}m_{Z}^2 \sin{2\beta}+\Delta m_{H_1}^2,\\
\label{MH1}
&&\xi_h = 1- \frac{(A_{X_1}^2)^2}{m_{R_1}^2 m_{H_1}^2}\,.
\end{eqnarray}
Here $A_{X_1}^2$ comes from the mixing of the neutral components of the Higgs doublets with the right-handed sneutrinos, and $m_{R_1}^2$ is the mass squared of the right-handed sneutrino, whose concrete expressions are given by
\begin{eqnarray}
&&A_{X_1}^2\simeq\sqrt{3}\lambda\upsilon\sin{2\beta}\left[2\upsilon_{\nu^{c}}\left({{3\lambda}\over{\sin{2\beta}}}-\kappa\right)-A_{\lambda}+{1\over2}(\Delta_{1R}+\Delta_{2R}) \right],\\
&&m_{R_1}^2= (A_\kappa+4\kappa\upsilon_{\nu^c})\kappa\upsilon_{\nu^c} +A_\lambda \lambda \upsilon_d \upsilon_u/\upsilon_{\nu^c} + \lambda^2 (2 \upsilon^2+3\Delta_{RR}),
\end{eqnarray}
where $\Delta_{1R}$, $\Delta_{2R}$ and $\Delta_{RR}$ are the radiative corrections~\cite{MASS}.
Comparing with the MSSM, $m_{H_1}^2$ in the $\mu\nu$SSM gets an additional term ${2\lambda_i \lambda_i s_{W}^2c_{W}^2\over e^2}m_{Z}^2 \sin{2\beta}$.  The radiative corrections $\bigtriangleup m_{H_{1}}^2$ can be computed more precisely by some public tools, for example, FeynHiggs~\cite{FeynHiggs-1,FeynHiggs-2,FeynHiggs-3,FeynHiggs-4,FeynHiggs-5,FeynHiggs-6,FeynHiggs-7,FeynHiggs-8}, SOFTSUSY~\cite{SOFTSUSY-1,SOFTSUSY-2,SOFTSUSY-3}, SPheno~\cite{SPheno-1,SPheno-2}, and so on. In the following numerical section, we will use the FeynHiggs-2.13.0 to calculate the radiative corrections for the Higgs boson mass about the MSSM part.

\section{The rare decay $h \rightarrow Z\gamma $ \label{sec3}}
\begin{figure}
\setlength{\unitlength}{1mm}
\centering
\begin{minipage}[c]{0.43\textwidth}
\includegraphics[width=2.5in]{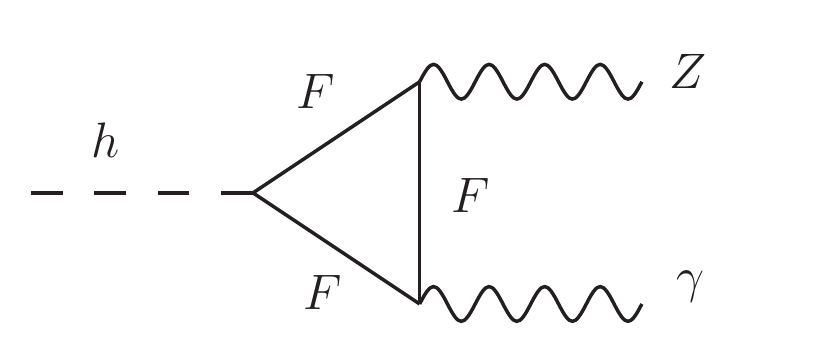}
\end{minipage}%
\begin{minipage}[c]{0.43\textwidth}
\includegraphics[width=2.5in]{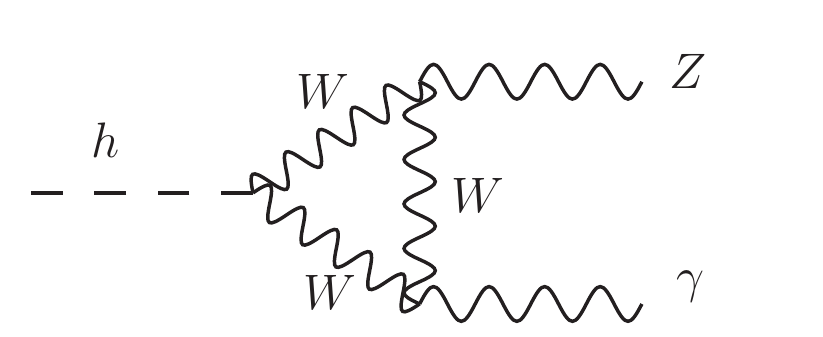}
\end{minipage}
\begin{minipage}[c]{0.43\textwidth}
\includegraphics[width=2.5in]{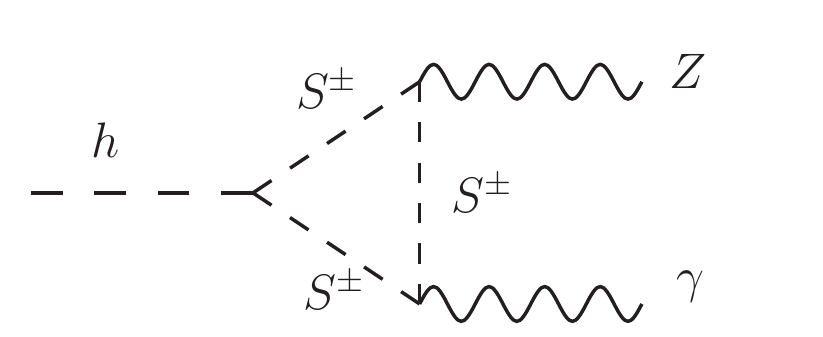}
\end{minipage}
\begin{minipage}[c]{0.43\textwidth}
\includegraphics[width=2.5in]{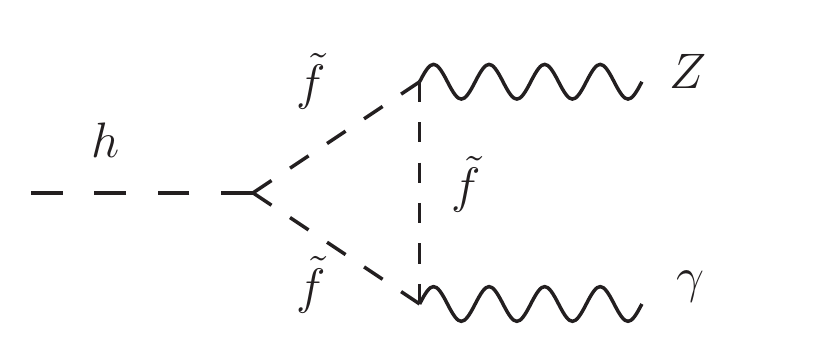}
\end{minipage}
\caption[]{The one-loop diagrams contributing to the decay $h\rightarrow Z\gamma$ in the $\mu\nu$SSM, where $F$ denotes the fermions and the charginos, $W$ is the $W$ gauge boson, $S^\pm$ denotes the charged scalars, and $\tilde{f}$ shows the sfermions. }
\label{one-loop}
\end{figure}

The $h\rightarrow Z\gamma$ coupling in the SM is similar to the $h\rightarrow \gamma\gamma$ coupling, which is built up by the heavy top quark and $W$ boson loops \cite{Higgs-SM}. In the supersymmetric models of the SM, there are more kinds of particles can make  contributions to the LO decay width, $W$ boson, the third-generation fermions ($f=t,b,\tau$) and the supersymmetric partners  \cite{Higgs}. In Fig.~\ref{one-loop}, we plot the one-loop diagrams contributing to the decay $h\rightarrow Z\gamma$ in the $\mu\nu$SSM, where $F$ denotes the fermions and the charginos, $W$ is the $W$ gauge boson, $S^\pm$ denotes the charged scalars, and $\tilde{f}$ shows the sfermions. Therefore, the decay width of the loop induced Higgs boson decay $h\rightarrow Z\gamma$ in the framework of the $\mu\nu$SSM can be mainly given as
\begin{eqnarray}
&&\hspace{-0.5cm}\Gamma_{{\rm{NP}}}(h\rightarrow Z\gamma)={\alpha G_{F}^2 m_{W}^2 m_{h}^3\over64\pi^4} {(1-{m_{Z}^2\over m_{h}^2}})^3
\Big|\sum\limits_{f=t,b,\tau} Q_{f}N_{f}\hat{v}_{f}g_{hff} A_{1/2}(x_f,\lambda_f)
+g_{{hWW}}A_{1}(x_W,\lambda_W)
\nonumber\\
&&\qquad\qquad\quad + (2c_W^2-1) g_{h S_\alpha^+ S_\alpha^-} {m_Z^2 \over m_{S_\alpha^\pm}^2}A_0(x_{S_\alpha^\pm},\lambda_{S_\alpha^\pm})
+\sum\limits_{\tilde{f}=U_I^+,D_I^-} N_c Q_{\tilde{f}}{\hat{v}_{\tilde{f}}} g_{h {\tilde{f}}{\tilde{f}}}{m_Z^2 \over m_{\tilde{f}}^2} A_0(x_{\tilde{f}},\lambda_{\tilde{f}})
\nonumber\\
&&\qquad\qquad\quad + \sum_{m,n=L,R} g_{h \chi_i \chi_i}^m g_{Z \chi_i \chi_i}^n  {2m_W \over m_{\chi_i}} A_{1/2}(x_{\chi_i},\lambda_{\chi_i})\Big|^2,
\end{eqnarray}
with $x_{i}={4m_{i}^2/ m_{h}^2}$, $\lambda_{i}={4m_{i}^2/ m_{Z}^2}$, $\hat{v}_{f}=(2I_f^3-4Q_f s_W^2)/ c_{W}$, $\hat{v}_{\tilde{f}_1}=(I_f^3\cos^2 \theta_f-Q_f s_W^2)/ c_{W}$, $\hat{v}_{\tilde{f}_2}=(I_f^3\sin^2 \theta_f-Q_f s_W^2)/ c_{W}$, $\theta_f$ is the mixing angle of sfermions $\tilde{f}_{1,2}$.  The form factors $A_0$, $A_{1/2}$ and $A_1$ are showed in Appendix~\ref{app1}. The concrete expressions of $g_{hff}$, $g_{{hWW}}$, $g_{h S_\alpha^+ S_\alpha^-}$, $g_{h {\tilde{f}}{\tilde{f}}}$ can be found in Ref~\cite{HZrr}. And the expressions of $g_{Z \chi_i \chi_i}^n $ and $g_{h \chi_i \chi_i}^n$ are
\begin{eqnarray}
&&g_{{Z\chi_i\chi_i}}^n=-{1\over { e} } C^{{Z\chi_i\bar{\chi}_i}}_n, \quad g_{{h\chi_i\chi_i}}^n=-{1\over { e} } C^{{S_1\chi_i\bar{\chi}_i}}_n \quad (n=L,R),
\end{eqnarray}
where $C^{{Z\chi_i\bar{\chi}_i}}_n$ and $C^{{S_1\chi_i\bar{\chi}_i}}_n$ ($h=S_1$) can be seen in Ref.~\cite{Zhang1}.

The decay width of $h\rightarrow Z\gamma$ at leading order (LO) in the $\mu\nu$SSM is mediated by charged heavy particle loops built up by $W$ bosons, standard fermions $f$, charged scalars $S^\pm_\alpha$, charginos $\chi_i$ and sfermions $\tilde{f}$.
When the supersymmetric particles are more heavy, the contributions of supersymmetric particles will be small. The signal strength of Higgs boson decay $h\rightarrow Z\gamma$ is a physical quantity that can be observed directly, and it can be written by
\begin{eqnarray}
&&\mu_{Z\gamma}^{{\rm{ggF}}}={ \sigma_{{\rm{NP}}}({\rm{ggF}})\over
\sigma_{{\rm{SM}}}({\rm{ggF}})}\:{{\rm{BR}}_{{\rm{NP}}}(h\rightarrow Z\gamma)\over
{\rm{BR}}_{{\rm{SM}}}(h\rightarrow Z\gamma)},
\label{eq-ratios}
\end{eqnarray}
normalized to the SM values, where ggF stands for gluon-gluon fusion.
One can evaluate the Higgs production cross sections
\begin{eqnarray}
&&{\sigma_{{\rm{NP}}}({\rm{ggF}})\over
\sigma_{{\rm{SM}}}({\rm{ggF}})} \approx {\Gamma_{{\rm{NP}}}(h\rightarrow gg) \over
\Gamma_{{\rm{SM}}}(h\rightarrow gg)} = {\Gamma_{{\rm{NP}}}^h\over
\Gamma_{{\rm{SM}}}^h}\: {\Gamma_{{\rm{NP}}}(h\rightarrow gg)/\Gamma_{{\rm{NP}}}^h \over
\Gamma_{{\rm{SM}}}(h\rightarrow gg)/\Gamma_{{\rm{SM}}}^h}\nonumber\\
&&\qquad\qquad\;\;={\Gamma_{{\rm{NP}}}^h\over
\Gamma_{{\rm{SM}}}^h}\: {{\rm{BR}}_{{\rm{NP}}}(h\rightarrow gg)\over
{\rm{BR}}_{{\rm{SM}}}(h\rightarrow gg)},
\label{eq-cross}
\end{eqnarray}
and the total decay width of the 125 GeV Higgs boson in the NP is~\cite{HZrr}
\begin{eqnarray}
&&\Gamma_{{\rm{NP}}}^h\simeq \sum\limits_{f=b,\tau,c,s} \Gamma_{{\rm{NP}}}(h\rightarrow f\bar{f})+ \sum\limits_{V=Z,W} \Gamma_{{\rm{NP}}}(h\rightarrow VV^*) \nonumber\\
&&\qquad\quad +\: \Gamma_{{\rm{NP}}}(h\rightarrow gg) +\Gamma_{{\rm{NP}}}(h\rightarrow \gamma\gamma) +\Gamma_{{\rm{NP}}}(h\rightarrow Z\gamma),
\label{eq-width}
\end{eqnarray}
where we neglected the little contribution which is rare or invisible, and the $\Gamma_{{\rm{SM}}}^h$ is the total decay width of the SM Higgs boson. Through Eqs.~(\ref{eq-ratios}-\ref{eq-width}), we can quantify the signal strength for the Higgs boson decay channel  $h\rightarrow Z\gamma$ in the $\mu\nu$SSM.

\section{Two-loop corrections of muon MDM\label{sec4}}

\begin{figure}
\setlength{\unitlength}{1mm}
\centering
\begin{minipage}[c]{1.0\textwidth}
\includegraphics[width=5.5in]{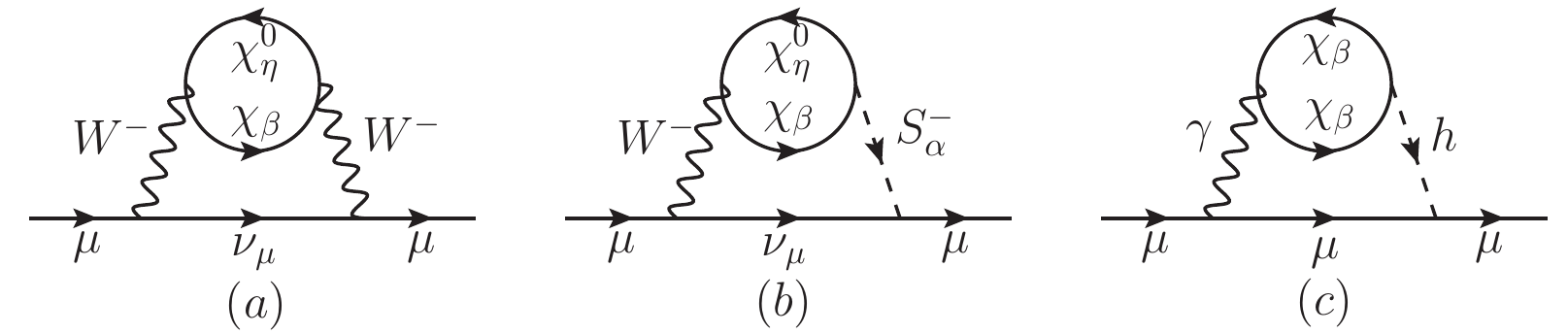}
\end{minipage}
\caption[]{The main two-loop rainbow diagram (a) and Barr-Zee type diagrams (b,c) in which a closed fermion loop is attached to the virtual gauge bosons or Higgs fields, the corresponding contributions to the muon MDM are obtained by attaching a photon on all possible ways to the internal particles.}
\label{Muon}
\end{figure}

The muon MDM in the $\mu\nu$SSM can be given as the effective Lagrangian
\begin{eqnarray}
&&\mathcal{L}_{MDM}={e\over 4m_{\mu}}a_{\mu}\overline{l}_{\mu}\sigma^{\alpha \beta}l_{\mu}F_{\alpha \beta},
\end{eqnarray}
where $l_\mu$ denotes the muon which is on-shell, $m_\mu$ is the mass of the muon, $\sigma^{\alpha\beta}=\frac{i}{2}[\gamma^\alpha,\gamma^\beta]$, $F_{\alpha\beta}$ represents the electromagnetic field strength and muon MDM, $a_\mu=\frac{1}{2}(g-2)_\mu$.
Including two-loop electroweak corrections, the muon MDM in the $\mu\nu$SSM can be written by
\begin{eqnarray}
&&a_{\mu}^{SUSY}=a_{\mu}^{one-loop}+a_{\mu}^{two-loop},
\end{eqnarray}
where the one-loop corrections $a_{\mu}^{one-loop}$ can be found in Ref.~\cite{muon}.

The two-loop diagrams can give important contributions to the muon MDM in a reasonable parameter space. According to Ref.~\cite{Yang:2009zzh,muon-a}, the main two-loop rainbow diagram (a) and Barr-Zee type diagrams (b,c) contributing to the muon MDM in the $\mu\nu$SSM are shown in Fig.~\ref{Muon}. Here, we ignore some two-loop diagrams which have low contributions, due to the decoupling theorem. In the $\mu\nu$SSM, the two-loop corrections are given as
\begin{eqnarray}
a_{\mu}^{two-loop}=a_{\mu}^{WW}+a_{\mu}^{WS}+a_{\mu}^{\gamma h},
\end{eqnarray}
where the terms $a_{\mu}^{WW},a_{\mu}^{WS},a_{\mu}^{\gamma h}$ are the contributions corresponding to Fig.~\ref{Muon} (a-c).
Under the assumption $m_{F}= m_{\chi_{\beta}} \simeq m_{{\chi}_{\eta}^0} \gg m_W$, the concrete expression can be approximately written as
\begin{eqnarray}
&&a_{\mu}^{WW}={G_F m_{\mu}^2 \over 192\sqrt{2}\pi^4} \{5(\big| C_{L}^{W \chi_{\beta} \overline{\chi}_{\eta}^0}\big| ^2
+\big| C_{R}^{W \chi_{\beta} \overline{\chi}_{\eta}^0}\big| ^2)-6(\big|C_{L}^{W \chi_{\beta} \overline{\chi}_{\eta}^0}\big| ^2-\big| C_{R}^{W \chi_{\beta} \overline{\chi}_{\eta}^0}\big| ^2)\nonumber\\
&&\;\;\quad\qquad+11\Re{(C_{L}^{W \chi_{\beta} \overline{\chi}_{\eta}^0}C_{R}^{W \chi_{\beta} \overline{\chi}_{\eta}^0 *})}\},\\
&&a_{\mu}^{WS}={G_{F} m_{\mu}m_{W}^2\Re{( C_{L}^{S_{\alpha}^{-}\chi_{9}^0 \overline{\chi}_4})} \over 128\pi^4m_F g_2}
\nonumber\\
&&\;\;\qquad\times\bigg\{\left[{179\over36}+{10\over3}J(m_{F}^2,m_{W}^2,m_{S_{\alpha}^{-}}^2)\right]\Re{(C_{L}^{W\chi_{\beta}\overline{\chi}_{\eta}^0}C_{L}^{W \chi_{\beta} \overline{\chi}_{\eta}^0}+C_{R}^{W \chi_{\beta}\overline{\chi}_{\eta}^0}C_{R}^{W\chi_{\beta}\overline{\chi}_{\eta}^0})}\nonumber\\
&&\;\;\qquad+\left[-{1\over 9}-{2 \over 3}J(m_{F}^2,m_{W}^2,m_{S_{\alpha}^{-}}^2)\right] \Re{(C_{L}^{W \chi_{\beta} \overline{\chi}_{\eta}^0}C_{R}^{W \chi_{\beta} \overline{\chi}_{\eta}^0}+C_{R}^{W \chi_{\beta}\overline{\chi}_{\eta}^0}C_{L}^{W\chi_{\beta}\overline{\chi}_{\eta}^0})}\nonumber\\
&&\;\;\qquad+\left[-{16\over 9}-{8\over3}J(m_{F}^2,m_{W}^2,m_{S_{\alpha}^{-}}^2)\right]\Re{(C_{L}^{W \chi_{\beta} \overline{\chi}_{\eta}^0}C_{L}^{W \chi_{\beta} \overline{\chi}_{\eta}^0}-C_{R}^{W \chi_{\beta}\overline{\chi}_{\eta}^0}C_{R}^{W\chi_{\beta}\overline{\chi}_{\eta}^0})}\nonumber\\
&&\;\;\qquad+\left[-{2\over 9}-{4\over 3}J(m_{F}^2,m_{W}^2,m_{S_{\alpha}^{-}}^2)\right]\Re{(C_{L}^{W \chi_{\beta} \overline{\chi}_{\eta}^0}C_{R}^{W \chi_{\beta} \overline{\chi}_{\eta}^0}-C_{R}^{W \chi_{\beta}\overline{\chi}_{\eta}^0}C_{L}^{W\chi_{\beta}\overline{\chi}_{\eta}^0})}
\bigg\},\\
&&a_{\mu}^{\gamma h}={-G_{F}m_{\mu}m_{W}^2 \over 32\pi^4 m_F}\left[1+\ln{m_{F}^2 \over m_{h}^2} \right]\Re{(C_{L}^{S_{1}\chi_{2} \overline{\chi}_{2}} C_{L}^{S_{1}\chi_{\beta} \overline{\chi}_{\beta}})},
\end{eqnarray}
where
\begin{eqnarray}
&&J(x,y,z)=\ln x-\frac{y\ln y-z\ln z}{y-z}.
\end{eqnarray}
Here, $\Re(\cdots)$ represents the operation to take the real part of a complex number, the concrete expressions for couplings $C$ can be found in Ref.~\cite{Zhang1}.

\section{Numerical analysis\label{sec5}}
In the $\mu\nu$SSM, there are many free parameters. We can take some appropriate parameter space, so that we can obtain a transparent numerical results. First, we make the minimal flavor violation (MFV) assumptions for some parameters, which assume
\begin{eqnarray}
&&\hspace{-0.9cm}{\kappa _{ijk}} = \kappa {\delta _{ij}}{\delta _{jk}}, \quad
{({A_\kappa }\kappa )_{ijk}} =
{A_\kappa }\kappa {\delta _{ij}}{\delta _{jk}}, \quad
\lambda _i = \lambda , \nonumber\\
&&\hspace{-0.9cm}
{({A_\lambda }\lambda )}_i = {A_\lambda }\lambda,\quad
{Y_{{e_{ij}}}} = {Y_{{e_i}}}{\delta _{ij}},\quad
{({A_e}{Y_e})_{ij}} = {A_{e}}{Y_{{e_i}}}{\delta _{ij}},\nonumber\\
&&\hspace{-0.9cm}
{Y_{{\nu _{ij}}}} = {Y_{{\nu _i}}}{\delta _{ij}},\quad
(A_\nu Y_\nu)_{ij}={a_{{\nu_i}}}{\delta _{ij}},\quad
m_{\tilde \nu_{ij}^c}^2 = m_{\tilde \nu_{i}^c}^2{\delta _{ij}}, \nonumber\\
&&\hspace{-0.9cm}m_{\tilde Q_{ij}}^2 = m_{{{\tilde Q_i}}}^2{\delta _{ij}}, \quad
m_{\tilde u_{ij}^c}^2 = m_{{{\tilde u_i}^c}}^2{\delta _{ij}}, \quad
m_{\tilde d_{ij}^c}^2 = m_{{{\tilde d_i}^c}}^2{\delta _{ij}}, \nonumber\\
&&\hspace{-0.9cm}m_{{{\tilde L}_{ij}}}^2 = m_{{\tilde L}}^2{\delta _{ij}}, \quad
m_{\tilde e_{ij}^c}^2 = m_{{{\tilde e}^c}}^2{\delta _{ij}}, \quad
\upsilon_{\nu_i^c}=\upsilon_{\nu^c},
\label{MFV}
\end{eqnarray}
where $i,\;j,\;k =1,\;2,\;3 $. $m_{\tilde \nu_i^c}^2$ can be constrained by the minimization conditions of the neutral scalar potential seen in Ref.~\cite{MASS}. To agree with experimental observations on quark mixing, one can have
\begin{eqnarray}
&&\hspace{-0.75cm}\;\,{Y_{{u _{ij}}}} = {Y_{{u _i}}}{V_{L_{ij}}^u},\quad
 (A_u Y_u)_{ij}={A_{u_i}}{Y_{{u_{ij}}}},\nonumber\\
&&\hspace{-0.75cm}\;\,{Y_{{d_{ij}}}} = {Y_{{d_i}}}{V_{L_{ij}}^d},\quad
(A_d Y_d)_{ij}={A_{d}}{Y_{{d_{ij}}}},
\end{eqnarray}
and $V=V_L^u V_L^{d\dag}$ denotes the CKM matrix.
\begin{eqnarray}
{Y_{{u_i}}} = \frac{{{m_{{u_i}}}}}{{{\upsilon_u}}},\qquad {Y_{{d_i}}} = \frac{{{m_{{d_i}}}}}{{{\upsilon_d}}},\qquad {Y_{{e_i}}} = \frac{{{m_{{l_i}}}}}{{{\upsilon_d}}},
\end{eqnarray}
where the $m_{u_{i}},m_{d_{i}}$ and $m_{l_{i}}$ stand for the up-quark, down-quark and charged lepton masses, and we can find the values of the masses from PDG~\cite{PDG1}.  Through our previous work~\cite{neu-mass6}, we have discussed in detail how the neutrino oscillation data constrain neutrino Yukawa couplings $Y_{\nu_i} \sim \mathcal{O}(10^{-7})$ and left-handed sneutrino VEVs $\upsilon_{\nu_i} \sim \mathcal{O}(10^{-4}\,{\rm{GeV}})$ in the $\mu\nu$SSM via the TeV scale seesaw mechanism.

Through analysis of the parameter space of the $\mu\nu$SSM in Ref.~\cite{mnSSM1}, we can take reasonable parameter values to  be $\lambda=0.1$, $\kappa=0.4$, $A_\lambda=500\;{\rm GeV}$, ${A_{\kappa}}=-300\;{\rm GeV}$ and $A_{u_{1,2}}=A_{d}=A_{e}=1\;{\rm TeV}$ for simplicity. Considering the direct search for supersymmetric particles~\cite{PDG1},  we take  $m_{{\tilde Q}_{1,2,3}}=m_{{\tilde u_{1,2}}^{c}}=m_{{\tilde d_{1,2,3}}^{c}}=2\;{\rm TeV}$, $m_{{\tilde L}}=m_{{{\tilde e}^c}}=1\;{\rm TeV}$, $M_3=2.5\;{\rm TeV}$. For simplicity, we will choose the gauginos' Majorana masses $M_1 = M_2$. As key parameters, $A_{u_{3}}=A_t$, $m_{{\tilde u}^c_3}$ and $\tan\beta$ greatly affect the lightest Higgs boson mass. Therefore, the free parameters that affect our next analysis are $\tan \beta ,\; \upsilon_{\nu^c},\; M_2,\; m_{{\tilde u}^c_3}$, and  $A_t$.

In the supersymmetric model, there is a close similarity between the anomalous magnetic dipole moment of muon and the branching ratio of $\bar{B}\rightarrow X_s\gamma$, in that both get large  $\tan\beta$ enhancements from a Higgsino-sfermion-fermion interaction vertex with a down-fermion Yukawa coupling~\cite{Martin:2001st}. So in the following, we also consider the constraint from  the branching ratio of $\bar{B}\rightarrow X_s\gamma$. The current combined  experimental data for the branching ratio of $\bar{B}\rightarrow X_s\gamma$ measured by CLEO~\cite{ref-CLEO}, BELLE \cite{ref-BELLE1,ref-BELLE2} and BABAR~\cite{ref-BABAR1,ref-BABAR2,ref-BABAR3} give~\cite{PDG1}
\begin{eqnarray}
{\rm{Br}}(\bar{B}\rightarrow X_s\gamma)=(3.49\pm0.19)\times10^{-4}.
\end{eqnarray}
In the next numerical analysis, we use our previous work about the rare decay $\bar{B}\rightarrow X_s\gamma$ in the $\mu\nu$SSM~\cite{ref-bsr}.

\subsection{Muon MDM}

\begin{table*}
\begin{tabular*}{\textwidth}{@{\extracolsep{\fill}}lllll@{}}
\hline
Parameters&Min&Max&Step\\
\hline
$\tan \beta$&4&40&2\\
$v_{\nu^{c}}/{\rm TeV}$&1&14&0.5\\
$m_{{\tilde u}^c_3}/{\rm TeV}$&1&4&0.3\\
$A_{t}/{\rm TeV}$&1&4&0.3\\
\hline
\end{tabular*}
\caption{Scanning parameters for the muon MDM with $M_2=\mu\equiv3\lambda \upsilon_{\nu^c}$.}
\label{tab1}
\end{table*}

Firstly, we analyze the muon MDM in the $\mu\nu$SSM. We define the physical quantity
\begin{eqnarray}
R_a\equiv{a_{\mu}^{two-loop}\over a_{\mu}^{one-loop}},
\end{eqnarray}
to show the ratio of two-loop corrections to one-loop corrections of the muon MDM.
To present numerical analysis, we scan the parameter space shown in Tab.~\ref{tab1}. Here the steps are large, because the running of the program is not very fast. However, the scanning parameter space is broad enough to contain the possibility of more. Considered that the light stop mass is easily ruled out by the experiment, we scan the parameter $m_{{\tilde u}^c_3}$ from 1 TeV.

\begin{figure}
\setlength{\unitlength}{1mm}
\centering
\begin{minipage}[c]{0.5\textwidth}
\includegraphics[width=2.8in]{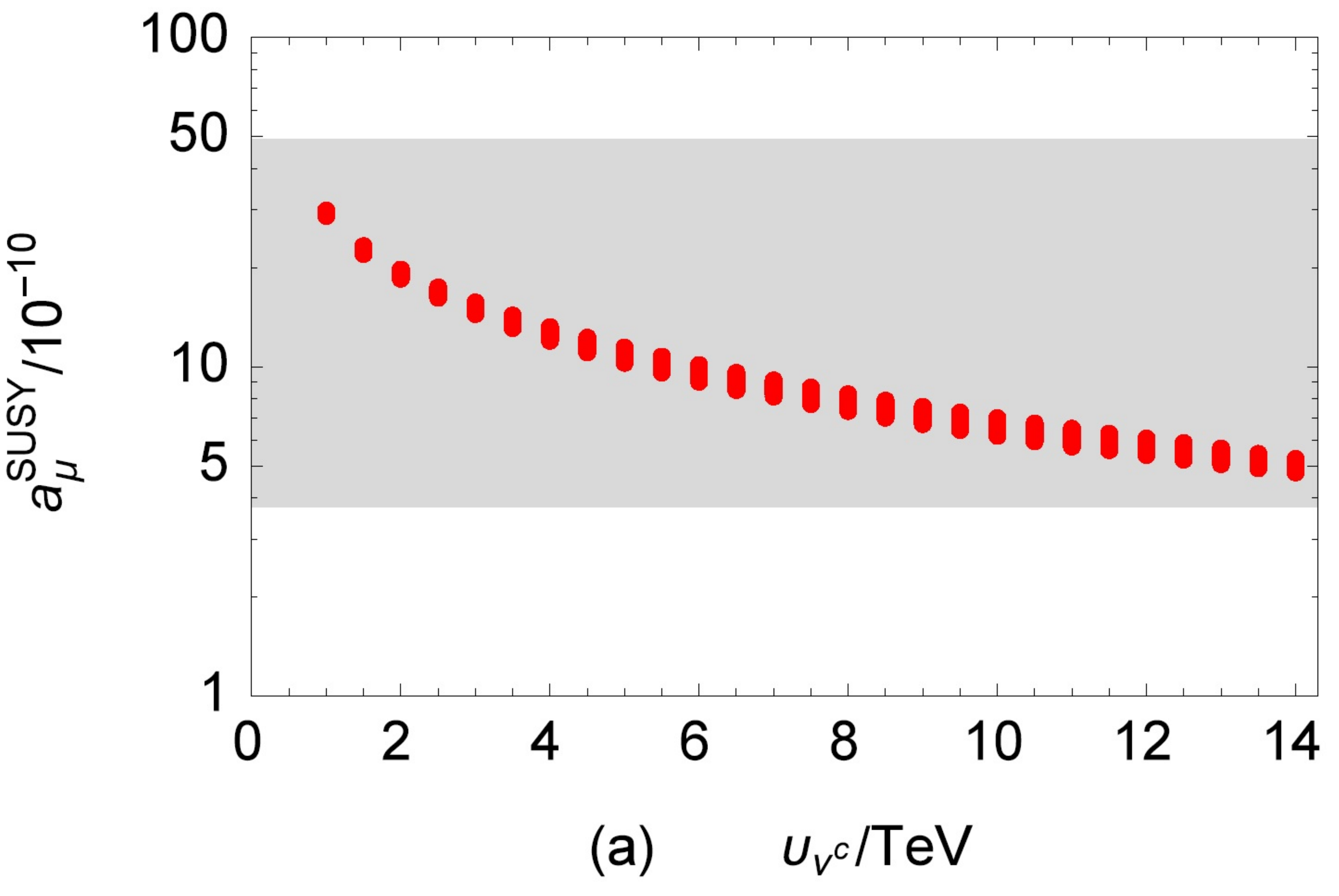}
\end{minipage}%
\begin{minipage}[c]{0.5\textwidth}
\includegraphics[width=2.8in]{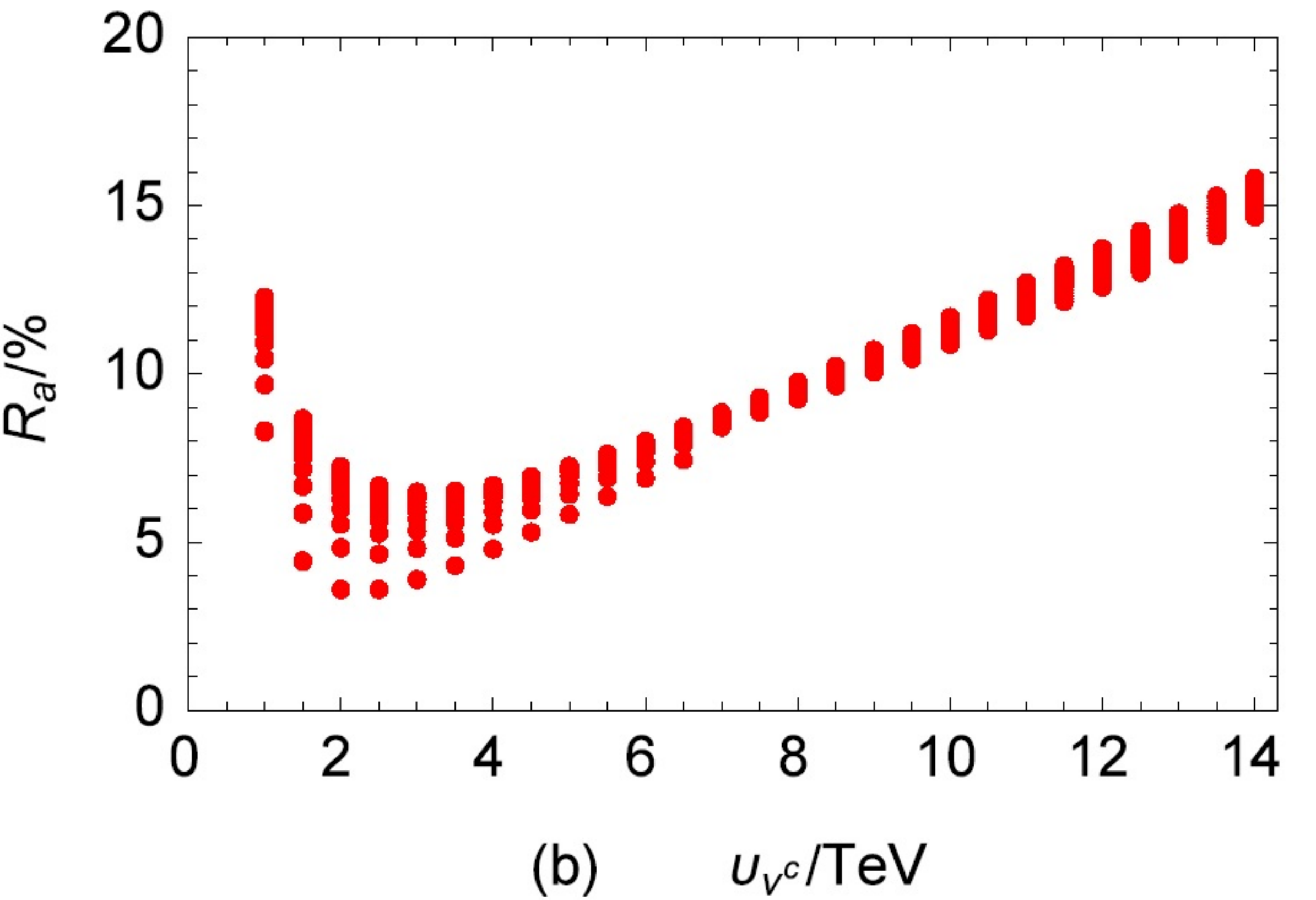}
\end{minipage}
\caption[]{$a_{\mu}^{SUSY}$ (a) and $R_a$ (b) vary with $v_{\nu^{c}}$, where the gray area denotes the muon MDM at $3.0\sigma$.}
\label{MuonR}
\end{figure}

\begin{figure}
\setlength{\unitlength}{1mm}
\centering
\begin{minipage}[c]{0.5\textwidth}
\includegraphics[width=2.8in]{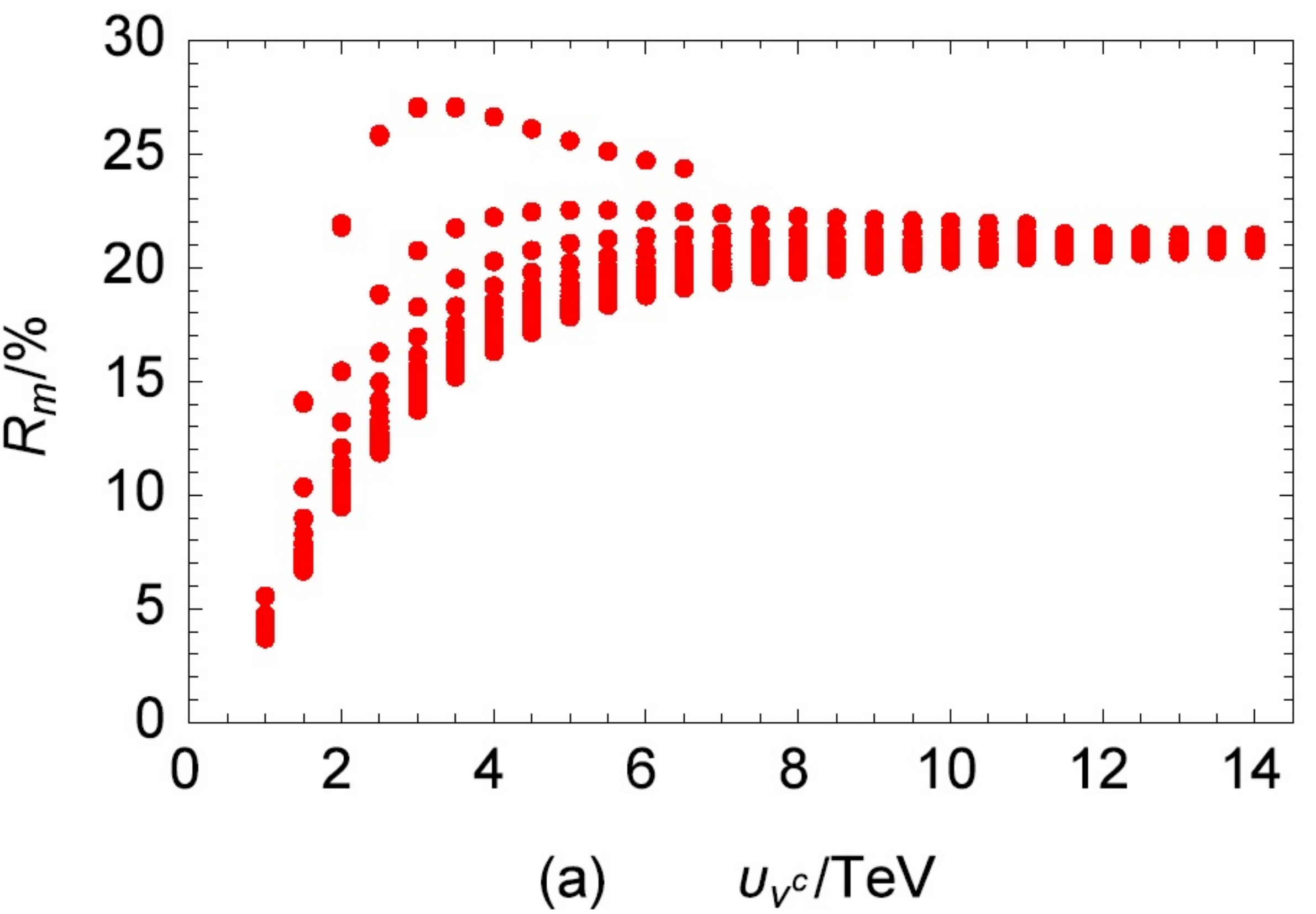}
\end{minipage}%
\begin{minipage}[c]{0.5\textwidth}
\includegraphics[width=2.8in]{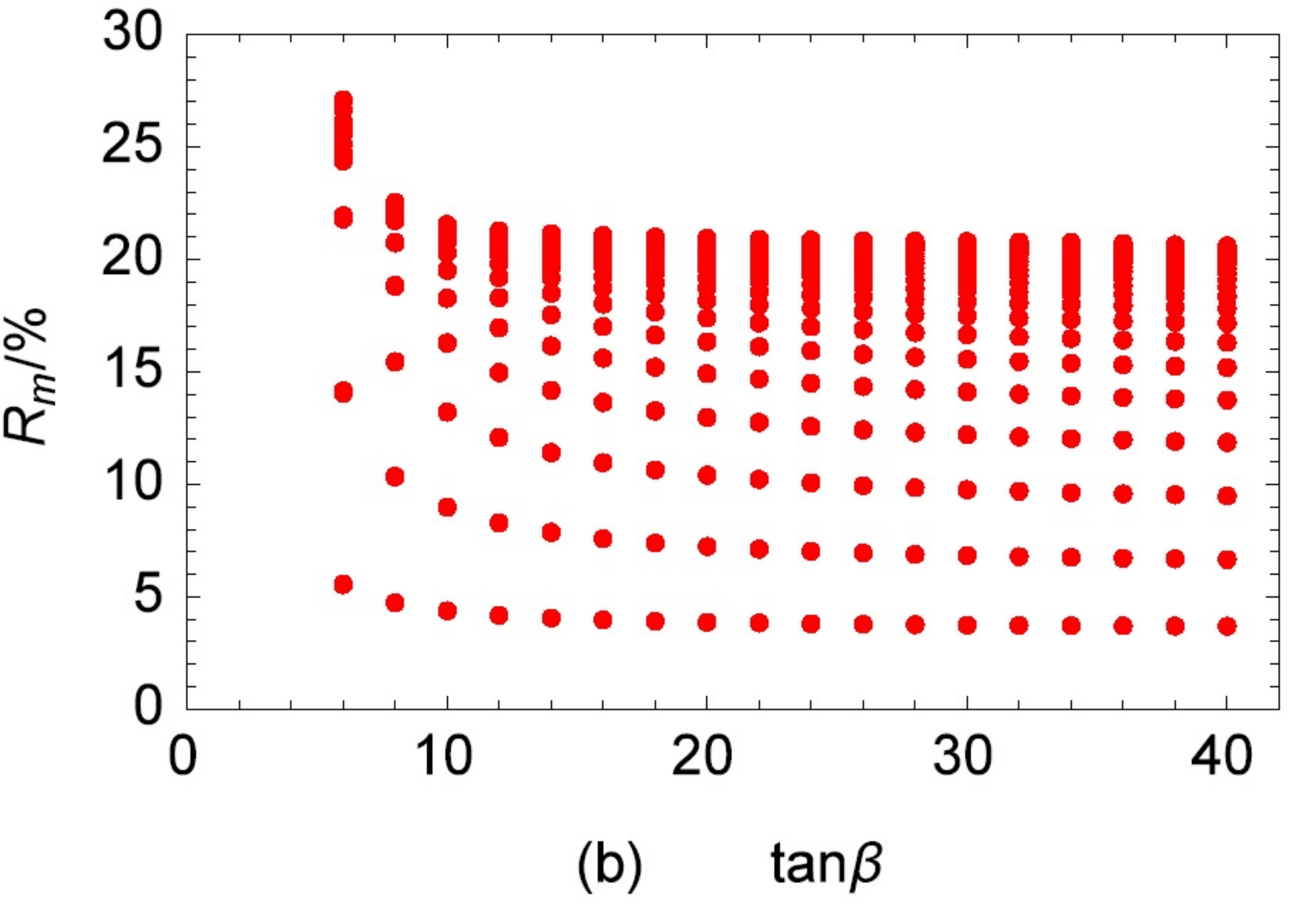}
\end{minipage}
\caption[]{$R_m$ varies with $\upsilon_{\nu^c}$ (a) and $\tan\beta$ (b).}
\label{MuonRm}
\end{figure}

Through scanning the parameter space in Tab.~\ref{tab1}, we plot Fig.~\ref{MuonR} and Fig.~\ref{MuonRm}, where the red dots are the corresponding physical quantity's values of the remaining parameters after being constrained by the lightest Higgs boson mass in the $\mu\nu$SSM with $124.68\,{\rm GeV}\leq m_{{h}} \leq125.52\:{\rm GeV}$, the branching ratio of $\bar{B}\rightarrow X_s\gamma$ in the $\mu\nu$SSM with $2.92\times 10^{-4} \leq {\rm{Br}}(\bar{B}\rightarrow X_s\gamma) \leq 4.06\times 10^{-4}$  and the muon anomalous MDM in the $\mu\nu$SSM with $3.7\times 10^{-10} \leq \Delta a_\mu \leq 49.9\times 10^{-10}$, where a $3 \sigma$ experimental error is considered.

In Fig.~\ref{MuonR}, we plot the muon anomalous MDM $a_{\mu}^{SUSY}$ and the ratio $R_a$ varying with the parameter $\upsilon_{\nu^{c}}$, where the gray area denotes the muon MDM at $3.0\sigma$ given in Eq.~(\ref{MDM-exp}). In Fig.~\ref{MuonR}(a), the numerical results show that the muon anomalous magnetic dipole moment $a_{\mu}^{SUSY}$ is decoupling with increasing $\upsilon_{\nu^{c}}$, which coincides with the decoupling theorem. We can see that the value of the muon anomalous MDM $a_{\mu}^{SUSY}$ in the $\mu\nu$SSM could reach the experimental  center value shown in Eq.~(\ref{MDM-exp}),  when $\upsilon_{\nu^{c}}$ is small.

To show the two-loop contributions of the muon MDM, Fig.~\ref{MuonR}(b) pictures the ratio $R_a$ varying with the parameter $\upsilon_{\nu^{c}}$. Normalized to the one-loop corrections of the muon MDM, the ratio $R_a$ can reach around 16\% when $\upsilon_{\nu^{c}}$ is large. Here, when $\upsilon_{\nu^{c}}$ is large, the one-loop corrections of the muon MDM are decoupling quickly than the two-loop corrections. The numerical results also show that the ratio $R_a$ can be about 12\% when $\upsilon_{\nu^{c}}$ is small. Therefore, the two-loop corrections also make important contributions to the muon anomalous MDM in the $\mu\nu$SSM.

To see the difference of two-loop contributions of muon MDM between the $\mu\nu$SSM and the MSSM, we define the physical quantity
\begin{eqnarray}
R_{m}\equiv {(a_{\mu}^{two-loop})_{\mu\nu\rm{SSM}}-(a_{\mu}^{two-loop})_{\rm{MSSM}} \over(a_{\mu}^{two-loop})_{\rm{MSSM}}}.
\end{eqnarray}
Here, $(a_{\mu}^{two-loop})_{\mu\nu\rm{SSM}}$ and $(a_{\mu}^{two-loop})_{\rm{MSSM}}$ respectively denote two-loop contributions of muon MDM of the $\mu\nu$SSM and those of the MSSM, which can be given in Sec.~\ref{sec4}.
In Fig.~\ref{MuonRm}, we show that $R_{m}$ varies with $\upsilon_{\nu^c}$ and $\tan\beta$.
In Fig.~\ref{MuonRm}(a), we can see that the ratio $R_{m}$ can reach about 27$\%$, when $\upsilon_{\nu^c}$ is around $3$ TeV. When the parameter $\upsilon_{\nu^c}$ is large, the maximum of the ratio $R_{m}$ is around 20$\%$. In Fig.~\ref{MuonRm}(b), we can know that when $\tan\beta$ is small, the ratio $R_{m}$ can be more large. Here, compared to the MSSM, the $\mu\nu$SSM has extra right-handed neutrinos which can give new contributions to the muon MDM. Simultaneously, the right-handed neutrino superfields lead to the mixing of right-handed neutrinos with the neutralinos.

\subsection{The decay $h \rightarrow Z\gamma $}

\begin{table*}
\begin{tabular*}{\textwidth}{@{\extracolsep{\fill}}lllll@{}}
\hline
Parameters&Min&Max&Step\\
\hline
$\tan \beta$&4&40&2\\
$v_{\nu^{c}}/{\rm TeV}$&1&14&0.5\\
$M_2/{\rm TeV}$&0.4&4&0.2\\
$m_{{\tilde u}^c_3}/{\rm TeV}$&1&4&0.3\\
$A_{t}/{\rm TeV}$&1&4&0.3\\
\hline
\end{tabular*}
\caption{Scanning parameters for the Higgs boson decay $h \rightarrow Z\gamma$.}
\label{tab2}
\end{table*}

\begin{figure}
\setlength{\unitlength}{1mm}
\centering
\begin{minipage}[c]{0.5\textwidth}
\includegraphics[width=2.8in]{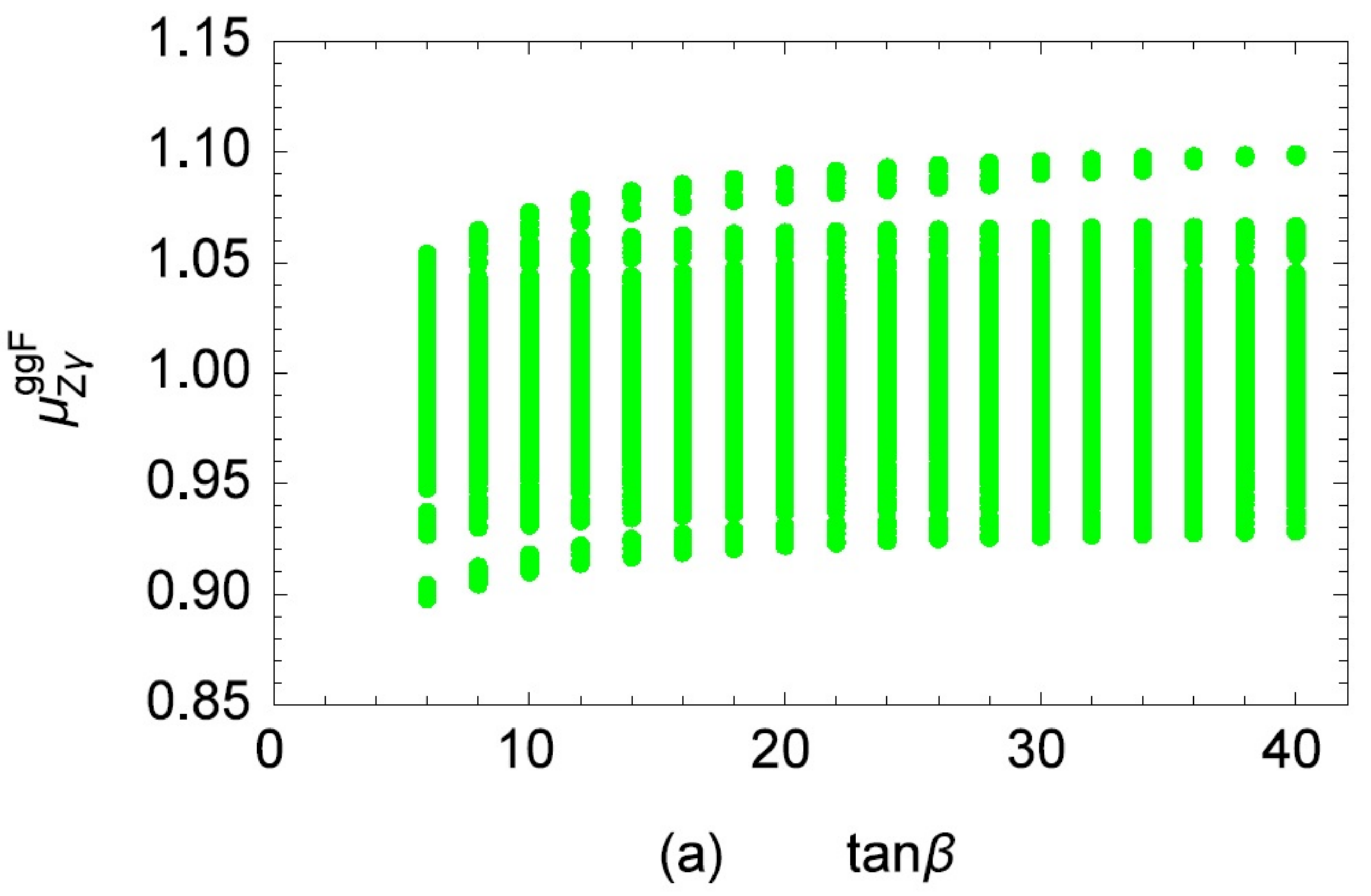}
\end{minipage}%
\begin{minipage}[c]{0.5\textwidth}
\includegraphics[width=2.8in]{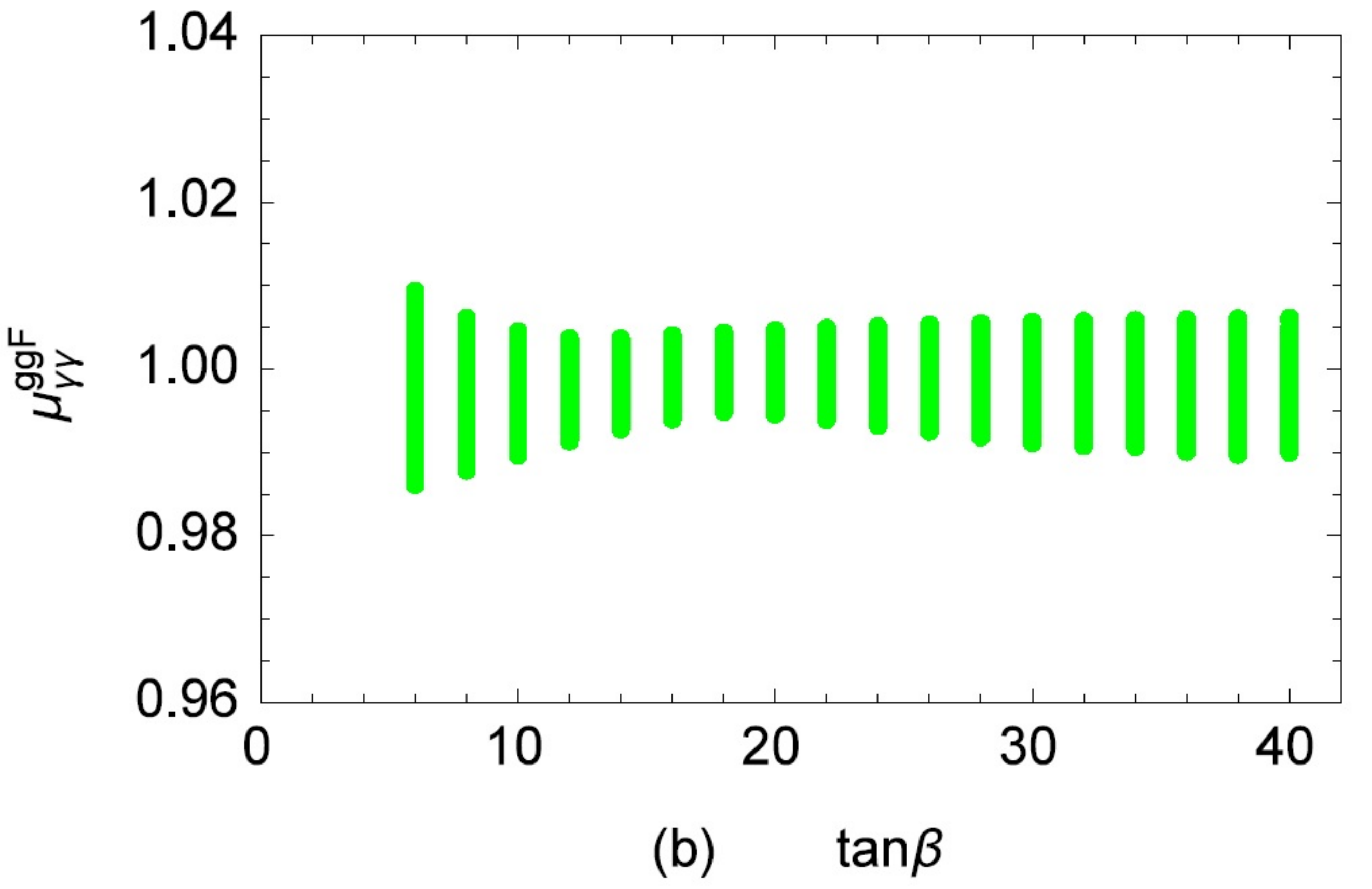}
\end{minipage}
\caption[]{The signal strength  $\mu_{Z\gamma}^{{\rm{ggF}}}$ (a) and  $\mu_{\gamma\gamma}^{{\rm{ggF}}}$ (b) versus $\tan \beta$.}
\label{RU}
\end{figure}

\begin{figure}
\setlength{\unitlength}{1mm}
\centering
\begin{minipage}[c]{0.5\textwidth}
\includegraphics[width=2.8in]{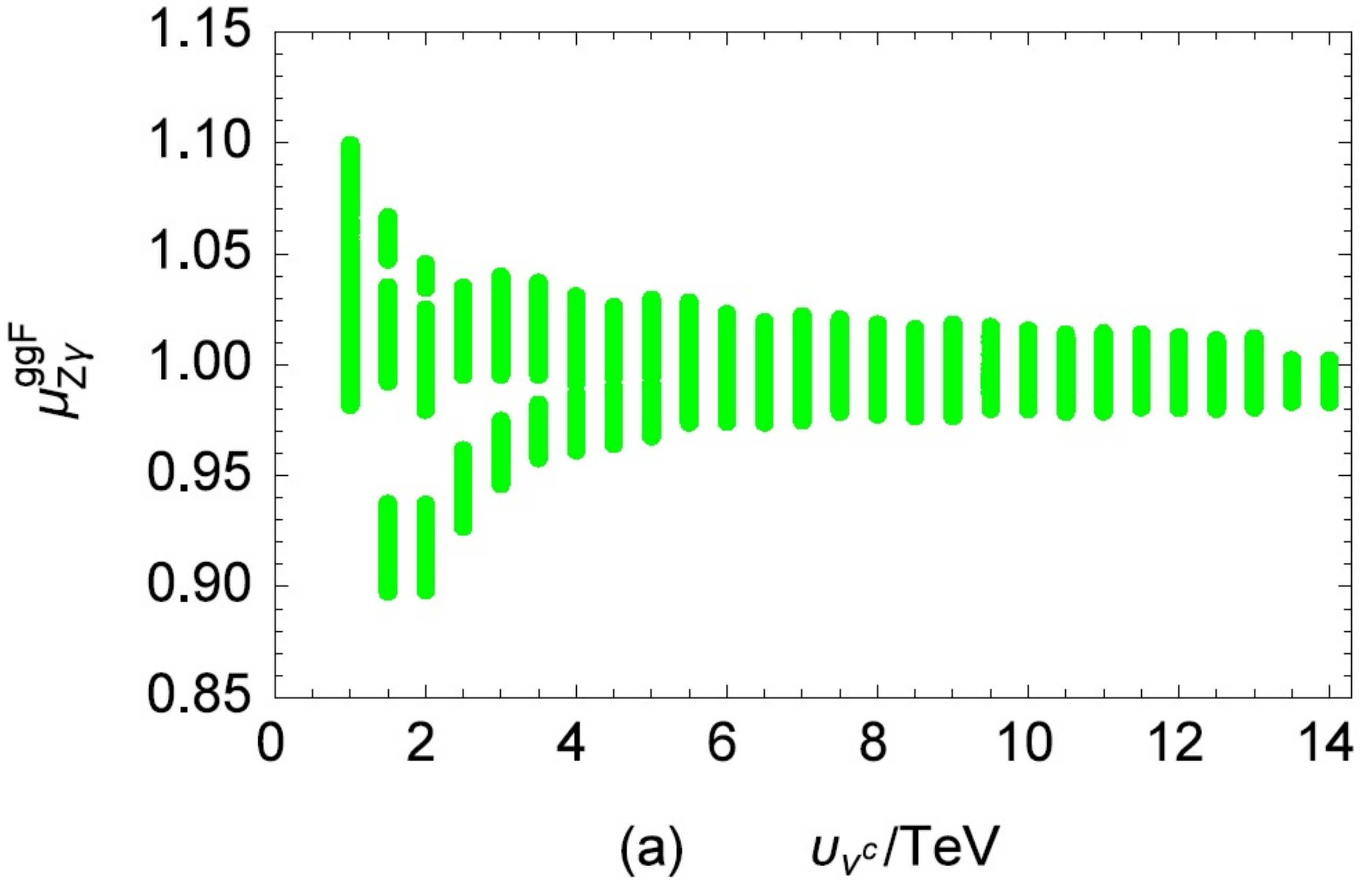}
\end{minipage}%
\begin{minipage}[c]{0.5\textwidth}
\includegraphics[width=2.8in]{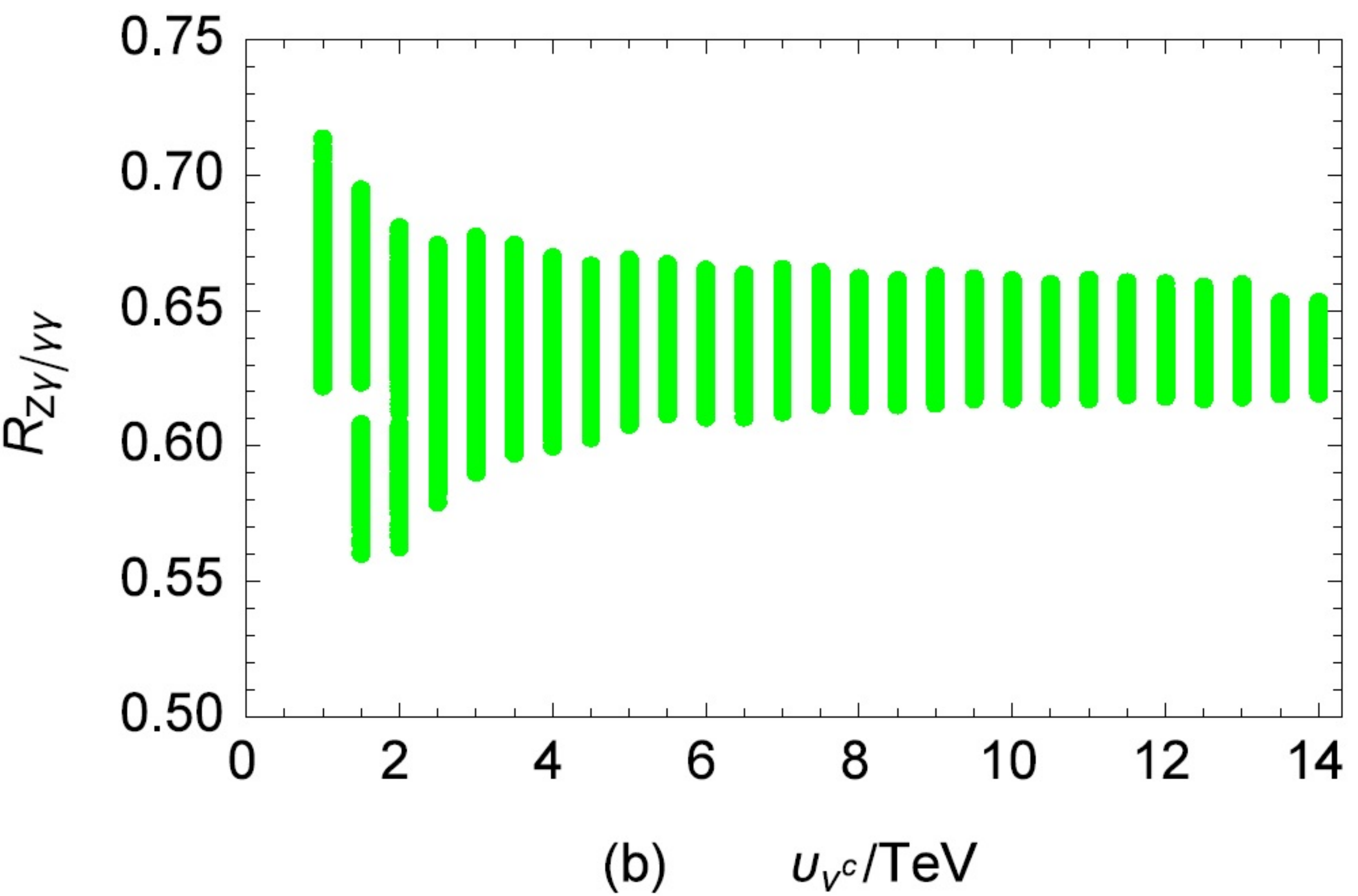}
\end{minipage}
\caption[]{The signal strength $\mu_{Z\gamma}^{{\rm{ggF}}}$ (a) and the ratio $R_{{Z\gamma / \gamma \gamma}}$ (b)  versus the parameter  $\upsilon_{\nu^{c}}$.}
\label{RZrrr}
\end{figure}

In this subsection, we present the numerical results of the signal strength for $h \rightarrow Z\gamma$. We plot Fig.~\ref{RU} and Fig.~\ref{RZrrr} through scanning the parameter space shown in Tab.~\ref{tab2}, where the green dots are the corresponding physical quantity's values of the remaining parameters after being constrained by the experimental constraints above. In Fig.~\ref{RU}(a), we plot the signal strength $\mu_{Z\gamma}^{{\rm{ggF}}}$ varying with $\tan \beta$.  The numerical results show that $0.9 \lesssim \mu_{Z\gamma}^{{\rm{ggF}}} \lesssim 1.1$. When $\tan \beta=6$, the signal strength $\mu_{Z\gamma}^{{\rm{ggF}}}$ can be down to 0.90 and up to 1.05. Here, the lightest Higgs boson in the $\mu\nu$SSM gets an additional term ${2\lambda_i \lambda_i s_{W}^2c_{W}^2\over e^2}m_{Z}^2 \sin{2\beta}$ in Eq.~(\ref{MH1}), comparing with the MSSM. Thus, the lightest Higgs boson in the $\mu\nu$SSM can easily account for the mass around 125 GeV, especially for small $\tan\beta$.

In Fig.~\ref{RU}(b), we also picture the signal strength $\mu_{\gamma\gamma}^{{\rm{ggF}}}$ varying with $\tan \beta$. We can see that the signal strength $\mu_{\gamma\gamma}^{{\rm{ggF}}}$ almost is around 1, which is consistent with the experimental value in the error range. Here, the relatively large stop mass and stau mass reduce the signal strength $\mu_{\gamma\gamma}^{{\rm{ggF}}}$. In Ref.~\cite{HZrr}, the signals of the Higgs boson decay channels $h\rightarrow\gamma\gamma$, $h\rightarrow VV^*$ ($V=Z,W$), and $h\rightarrow f\bar{f}$ ($f=b,\tau$) in the $\mu\nu$SSM have been investigated. When the lightest stop mass $m_{{\tilde t}_1}\gtrsim 700\;{\rm GeV}$ and the lightest stau mass $m_{{\tilde \tau}_1}\gtrsim 300\;{\rm GeV}$, the signal strengths of these Higgs boson decay channels  in the $\mu\nu$SSM are in agreement with those in the SM.

Through Fig.~\ref{RU}(a) and Fig.~\ref{RU}(b), the numerical results show that the signal strength $\mu_{Z\gamma}^{{\rm{ggF}}}$ in the $\mu\nu$SSM still has a large deviation from 1, even though the signal strength $\mu_{\gamma\gamma}^{{\rm{ggF}}}$ in the $\mu\nu$SSM is in keeping with that in the SM.

We plot the signal strength $\mu_{Z\gamma}^{{\rm{ggF}}}$ versus the parameter $\upsilon_{\nu^{c}}$ in Fig.~\ref{RZrrr}(a). The numerical results present that the signal strength $\mu_{Z\gamma}^{{\rm{ggF}}}$ can have a large deviation from 1, when the value of the parameter $\upsilon_{\nu^{c}}$ is small. The parameter $\upsilon_{\nu^{c}}$ directly affects the mass of chargino. The small chargino mass gives a large contribution to the signal strength $\mu_{Z\gamma}^{{\rm{ggF}}}$. In addition, the parameter $\upsilon_{\nu^{c}}$ leads to the mixing of the neutral components of the Higgs doublets with the sneutrinos. The mixing affects the lightest Higgs boson mass and the Higgs couplings, which is different from the MSSM.

To see more clearly, we also plot the ratio $R_{{Z\gamma/\gamma\gamma}} \equiv \Gamma_{{\rm{NP}}}(h\rightarrow Z\gamma) / \Gamma_{{\rm{NP}}}(h\rightarrow \gamma\gamma)$ versus the parameter $\upsilon_{\nu^{c}}$ in Fig.~\ref{RZrrr}(b). We can see that $0.55 \lesssim R_{{Z\gamma/\gamma\gamma}}  \lesssim 0.71$, when $\upsilon_{\nu^{c}}$ is small. Here, small value of the parameter $\upsilon_{\nu^{c}}$ can give more large contributions to the decay width $\Gamma_{{\rm{NP}}}(h\rightarrow Z\gamma)$ than $\Gamma_{{\rm{NP}}}(h\rightarrow \gamma\gamma)$. Thus, the signal strength $\mu_{Z\gamma}^{{\rm{ggF}}}$ in the $\mu\nu$SSM has a large deviation from 1, through small value of the parameter $\upsilon_{\nu^{c}}$ which affects the mass of chargino and leads to the mixing of the neutral components of the Higgs doublets with the sneutrinos.

\section{Summary\label{sec6}}

In the framework of the $\mu\nu$SSM, the three singlet right-handed neutrino superfields $\hat{\nu}_i^c$ are introduced to solve the $\mu$ problem of the MSSM and generate three tiny Majorana neutrino masses at the tree level through the seesaw mechanism. The gravitino or the axino in  the $\mu\nu$SSM also can be a dark mater candidate. The right-handed sneutrino VEVs lead to the mixing of the neutral components of the Higgs doublets with the sneutrinos. Therefore, the mixing would affect the lightest Higgs boson mass and the Higgs couplings, which gives a rich phenomenology in the Higgs sector of  the $\mu\nu$SSM, being different from the MSSM.

In this paper, we analyze the signal strength of the Higgs boson decay $h\rightarrow Z\gamma$ in the $\mu\nu$SSM. Even though the signal strength of  $h\rightarrow \gamma\gamma$ in the $\mu\nu$SSM is in accord with that in the SM, the signal strength of $h\rightarrow Z\gamma$ in the $\mu\nu$SSM still has a large deviation from 1, due to the small mass of chargino and the mixing of the neutral components of the Higgs doublets with the sneutrinos. The present observed 95\% CL upper limit on the signal strength of the $h\rightarrow Z\gamma$ decay still is 6.6~\cite{hZr-ATLAS2}. However, high luminosity or high energy large collider~\cite{ref-100pp,ref-HL,ref-CEPC} built in the future will detect the Higgs boson decay $h\rightarrow Z\gamma$, which may see the indication of new physics.

Here, we also consider the two-loop corrections of the muon anomalous MDM in the $\mu\nu$SSM. Normalized to the one-loop corrections of the muon MDM, the two-loop corrections in the $\mu\nu$SSM can be around 16\%. Compared to the MSSM, the $\mu\nu$SSM has extra right-handed neutrinos which can give new contributions to the muon anomalous MDM. Therefore, the two-loop corrections also make important contributions to the muon anomalous MDM in the $\mu\nu$SSM. In near future, the Muon g-2 experiment E989 at Fermilab~\cite{ref-muon-exp,ref-muon-exp1} will measure the muon anomalous magnetic dipole moment with unprecedented precision, which may reach a 5$\sigma$ deviation from the SM, constituting an augury for new physics beyond the SM.

\begin{acknowledgments}
\indent\indent
The work has been supported by the National Natural Science Foundation of China (NNSFC) with Grants No. 11705045, No. 11647120, No. 11535002, the youth top-notch talent support program of the Hebei Province,  and Midwest Universities Comprehensive Strength Promotion project.
\end{acknowledgments}

\appendix

\section{FORM FACTORS\label{app1}}

\begin{eqnarray}
&&A_{1/2}(\tau, \lambda)=I_1(\tau, \lambda)-I_{2}(\tau, \lambda), \\
&&A_{1}(\tau, \lambda)=c_{W}\Big\{{4\Big(3-{s_{W}^2 \over c_{W}^2}\Big)I_{2}(\tau, \lambda)+\Big[\Big(1+{2 \over \tau}\Big){s_{W}^2 \over c_{W}^2}-\Big(5+{2\over \tau}\Big)\Big]I_1(\tau, \lambda)}\Big\}, \\
&&A_0(\tau, \lambda)=I_1(\tau, \lambda),\\
&&I_1(\tau, \lambda)= \frac{\tau\lambda}{2(\tau-\lambda)} + \frac{\tau^2\lambda^2}{2(\tau-\lambda)^2} \Big[f(\tau^{-1}) -  f(\lambda^{-1})\Big]  + \frac{\tau^2\lambda}{(\tau-\lambda)^2} \Big[g(\tau^{-1}) -  g(\lambda^{-1})\Big]  ,\quad\\
&&I_2(\tau, \lambda)=-\frac{\tau\lambda}{2(\tau-\lambda)}\Big[f(\tau^{-1})-  f(\lambda^{-1})\Big],\\
&&f(\tau)=\left\{\begin{array}{l}\arcsin^2\sqrt{\tau},  \hspace{2.6cm} \tau\le1;   \\
-{1\over4}\Big[\log{1+\sqrt{1-1/\tau}\over1-\sqrt{1-1/\tau}}-i\pi\Big]^2,  \quad \tau>1,  \end{array}\right.\\
&&g(\tau)=\left\{\begin{array}{l}\sqrt{\tau^{-1}-1}\, {\arcsin\sqrt{\tau}},  \hspace{1.95cm} \tau\geq1;   \\
{{\sqrt{1-\tau^{-1}}}\over2}\Big[\log{1+\sqrt{1-1/\tau}\over1-\sqrt{1-1/\tau}}-i\pi\Big],  \quad \tau<1.  \end{array}\right.
\end{eqnarray}

\end{document}